Formation of organic hazes in CO$_2$-rich sub-Neptune atmospheres within the graphite-stability regime

Sai Wang[1#], Zhengbo Yang[1#], Chao He[1*], Haixin Li[1], Yu Liu[1], Yingjian Wang[1], Xiao'ou Luo[1], Sarah E. Moran[2], Cara Pesciotta[3], Sarah M. Hörst[3,4], Julianne I. Moses[5], Véronique Vuitton[6], Laurène Flandinet[6]

1. School of Earth and Space Sciences, University of Science and Technology of China, Hefei 230026, China
2. NHFP Sagan Fellow, NASA Goddard Space Flight Center, Greenbelt, MD 20771, USA.
3. Department of Earth and Planetary Sciences, Johns Hopkins University, Baltimore, MD, USA.
4. Space Telescope Science Institute, Baltimore, MD, USA.
5. Space Science Institute, Boulder, CO, USA.
6. Univ. Grenoble Alpes, CNRS, IPAG, Grenoble, France.

#These authors contributed equally to this work.
*Corresponding Author: Chao He (chaohe23@ustc.edu.cn)

**Abstract**

Super-Earths and sub-Neptunes are the most common exoplanets, with a "radius valley" suggesting that super-Earths may form by shedding sub-Neptunes' gaseous envelopes. Exoplanets that lie closer to the super-Earth side of the valley are more likely to have lost a significant fraction of their original H/He envelopes and become enriched in heavier elements with CO$_2$ gaining in abundance. It remains unclear which types of haze would form in such atmospheres, potentially significantly affecting spectroscopic observations. To investigate this, we performed laboratory simulations of two CO$_2$-rich gas mixtures (with 2000 times solar metallicity at 300 K and 500 K). We found that under plasma irradiation, organic hazes were produced at both temperatures with higher haze production rate at 300 K probably because condensation occurs more readily at lower temperature. Gas-phase analysis demonstrates the formation of various hydrocarbons, oxygen- and nitrogen-containing species, including reactive gas precursors like C$_2$H$_4$, CH$_2$O, and HCN, for haze formation. The compositional analysis of the haze particles reveals various functional groups and molecular formulas in both samples. The 500 K haze sample has larger average molecular sizes, higher degree of unsaturation with more double or triple bonds presence, and higher nitrogen content incorporated as N-H, C=N bonds, indicating different haze formation pathways. These findings not only improve the haze formation theories in CO$_2$-rich exoplanet atmospheres but also offer important implications for the interpretation of future observational data.

**Keywords:** Exoplanet atmospheric chemistry; Photochemical haze; Organic compounds; Laboratory simulations; Graphite haze

# 1. Introduction

In the last two decades, we have discovered nearly 6000 exoplanets, and found that planets smaller than Neptune but larger than Earth are most common in our galaxy (N. M. Batalha et al. 2013; F. Fressin et al. 2013; A. W. Howard et al. 2012; M. Mayor et al. 2011). The Kepler data and radial velocity (RV) data have revealed a bimodal distribution of planet radii with one peak at ~1.3 Earth radii ($R_\oplus$), the other at ~2.4 $R_\oplus$, and a minimum between them at ~1.75 $R_\oplus$, commonly named the "radius valley" (B. J. Fulton et al. 2017). The two peaks correspond to smaller planets (likely rocky) and larger intermediate-size planets, respectively. Those exoplanets with sizes less than 1.6 $R_\oplus$ are called super-Earths, and those with sizes of 1.8 $R_\oplus$ or larger are called sub-Neptunes. They are the two most populous sizes of exoplanets and have no analogues in the present-day Solar System. The formation and evolution processes of these exoplanets remain uncertain.

There are few planets found with radii in between these two populations, although the exact location and properties defining this radius valley may depend on stellar type (T. A. Berger et al. 2018; A. Bonfanti et al. 2024; B. J. Fulton & E. A. Petigura 2018; R. Luque & E. Pallé 2022; V. Van Eylen et al. 2018). This radius valley is interpreted as evidence that super-Earths might form by shedding sub-Neptunes' gaseous envelopes (R. Cloutier & K. Menou 2020; A. Gupta & H. E. Schlichting 2019, 2020; D. Modirrousta-Galian et al. 2020; C. Mordasini 2020; J. G. Rogers & J. E. Owen 2021). Two main physical mechanisms have been proposed to explain the existence of the radius valley: the photoevaporation (S. Jin et al. 2014; E. D. Lopez & J. J. Fortney 2013; C. Mordasini 2020; J. E. Owen 2019; J. E. Owen & Y. Wu 2013; J. G. Rogers & J. E. Owen 2021) and core-powered mass loss mechanisms (R. Cloutier & K. Menou 2020; A. Gupta & H. E. Schlichting 2019, 2020; D. Modirrousta-Galian et al. 2020). Prior to the discovery of the radius valley with observations, planetary formation/evolution models predicted a valley in the distribution of planet radii caused by the photoevaporation of planetary envelopes by the X-ray and extreme ultraviolet irradiation from the host star (J. E. Owen & Y. Wu 2013). The observed radius gap is consistent with the "photoevaporation valley". Motivated by its observational discovery, a variety of subsequent, more detailed models have quantified the photoevaporation scenario further (D. Modirrousta-Galian et al. 2020; C. Mordasini 2020). The photoevaporation valley defines the boundary between planets with a mass large enough to hold on to their gas envelope, and planets that have been stripped of their atmospheres and consist of the remnant core. An alternative mechanism, core-powered mass-loss, is unrelated to the high-energy flux and yet also naturally reproduces the observed bimodal distribution (H. Chen & L. A. Rogers 2016; A. Gupta & H. E. Schlichting 2019; J. G. Rogers & J. E. Owen 2021). In this scenario, the luminosity of the cooling rocky core, which can completely erode light envelopes while preserving heavy ones, also produces a gap within the intermediate-sized planet population. Future observations are expected to help distinguish between two different scenarios. For instance, the photoevaporation and core-powered mass loss mechanisms predict a different dependence of the radius valley on the system age (S. Ginzburg et al. 2018; D. Modirrousta-Galian et al. 2020; J. G. Rogers & J. E. Owen 2021). The mechanism responsible for the radius valley is debated, but it is very likely both mass-loss mechanisms take place to some extent and contribute to the observed radius valley, while their relative contributions vary among different exoplanets (T. A. Berger et al. 2020; J. E. Owen & H. E. Schlichting 2024; J. G. Rogers & J. E. Owen 2021).

Regardless of the responsible mechanism, planets closer to the gap are more likely to have lost a significant fraction of their original H/He envelopes. Many of these 1.5 to 3.0 $R_\oplus$ exoplanets start out as $H_2$-rich sub-Neptunes, but the smaller planets lose their original atmospheres due to either photoevaporation or core-powered mass loss processes. As hydrogen escapes, these atmospheres become more enriched in heavy elements (i.e., more "metallic"), with $CO_2$ gaining in abundance. As $CO_2$ becomes dominant, the atmosphere enters a regime where graphite is thermodynamically stable, with graphite condensation anticipated, as indicated by equilibrium models (R. Mbarek & E. M.-R. Kempton 2016; J. I. Moses et al. 2013). The formed graphite hazes could impact spectroscopic observations with JWST, the Atmospheric Remote-sensing Infrared Exoplanet Large-survey (ARIEL) mission, and the Extremely Large Telescopes (ELTs). A handful of sub-Neptunes and Neptune-sized exoplanets near the radius valley, such as GJ 436 b (S. Mukherjee et al. 2025) , GJ 1214 b (K. Ohno et al. 2025), GJ 3470 b (T. G. Beatty et al. 2024), L98-59 d (A. Gressier et al. 2024) , TOI-776 c (J. Teske et al. 2025) , TOI-421 b (B. Davenport et al. 2025) , and TOI-270 d (J. Yang & R. Hu 2024), could be in the graphite-stability regime. These exoplanets have been observed by JWST, with their spectra showing relatively flat characteristics. This suggests that the atmospheres of these exoplanets might be dominated by haze, although the nature of the haze is unknown. More sub-Neptunes, including TOI-1130 b, TOI-178 d, g, TOI 1801 b, LP 791-18 c and HD 86226 c, have been selected as observational targets in the JWST Guaranteed Time Observations (GTO) and General Observations (GO) programs. Therefore, it is essential to focus on the atmospheric composition of these exoplanets in order to better interpret the data obtained from JWST.

Although graphite formation and condensation are expected over a wide phase space, the chemical kinetics involved with its formation are unclear. Graphite production is typically an energy-intensive process that requires extremely high temperatures—for example, the Acheson process operates at temperatures exceeding 3000 K (E. G. Acheson 1896). Despite its thermodynamic favorability, it remains uncertain whether graphite can form in the upper regions of $CO_2$-rich atmospheres, where it would be observable by telescopes such as JWST. Here we present a laboratory study of the chemical processes in sub-Neptune atmospheres within the graphite-stability regime. We first determine whether graphite hazes form under relevant conditions, investigate the composition of the gas and solid products, and then discuss the potential chemical processes and their impact on the observations of sub-Neptune atmospheres.

## 2. Materials and Experimental

### 2.1. Experimental Setup and Procedure

The experiments were conducted using the Planetary Haze Research (PHAZER) setup (Figure 1), which is specifically designed to investigate a range of planetary atmospheres. This facility can be operated at a broad range of temperatures and pressures using a variety of reactant gases (C. He et al. 2017).

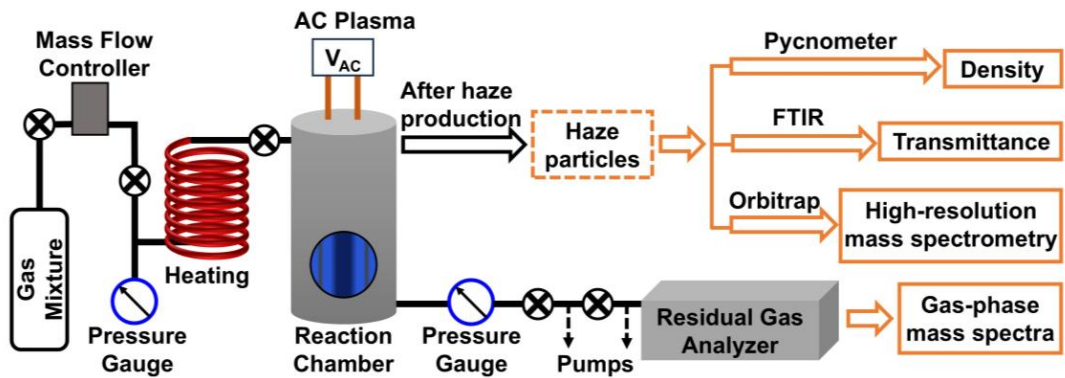

**Figure 1.** Simplified schematic of the experimental set-up and the analytical method for the current study.

### 2.1.1. Gas Mixtures and Conditions

Theoretical modeling suggests wide variety of possible atmospheres in sub-Neptunes. Many of the predictions of the atmospheric composition of exoplanets are grounded in Solar System science (P. Gao et al. 2021; E. J. Lee & N. J. Connors 2021; E. Sciamma-O'Brien et al. 2014; K. B. Stevenson et al. 2010; P. Woitke et al. 2021; E. T. Wolf 2017). We focused on atmospheres that fall within the graphite-stability regime in order to explore whether graphite haze does indeed form in sub-Neptunes. We used thermochemical equilibrium calculations to guide the choice of our starting atmospheric compositions (Figure 2) (J. I. Moses et al. 2013). To represent high-metallicity atmospheres appropriate for sub-Neptune exoplanets, we considered 2000× solar metallicity for our initial gas-phase composition. This choice of metallicity (2000× solar metallicity) was motivated by the observation that some exoplanets, particularly sub-Neptunes like GJ-1214 b (E. M.-R. Kempton et al. 2023; I. Malsky et al. 2025; K. Ohno et al. 2025; E. Schlawin et al. 2024), may experience significant atmospheric metal enrichment during their formation, especially after undergoing processes such as photoevaporation. In these atmospheres, $CO_2$ may become the dominant gas and thus lead to graphite hazes but it's also possible for sub-Neptunes to have $CH_4$-rich atmospheres (D. J. Bower et al. 2025) which would lead to photochemical hazes.

The equilibrium compositions (J. I. Moses et al. 2013) were calculated for two representative temperatures of 300 K and 500 K at 3 mbar. We performed the calculation with and without graphite in the list of species. To maintain a manageable level of experimental complexity, only gases with a calculated abundance of 1% or higher are included in the resulting experiments. We started the experiments with the gas compositions that contain no graphite on the left in Figure 2 to see whether graphite would form as predicted from the thermochemical-equilibrium calculations with graphite included (right in Figure 2). For the 300 K condition, the calculated starting gas mixture includes 43% $CO_2$, 19% $H_2O$, 18% $CH_4$, 9% $N_2$, and 11% He. For the 500 K condition, the gas composition slightly changes with the presence of $H_2$ at 6%. More details about the chemical equilibrium model can be found in previous studies (C. He et al. 2022).

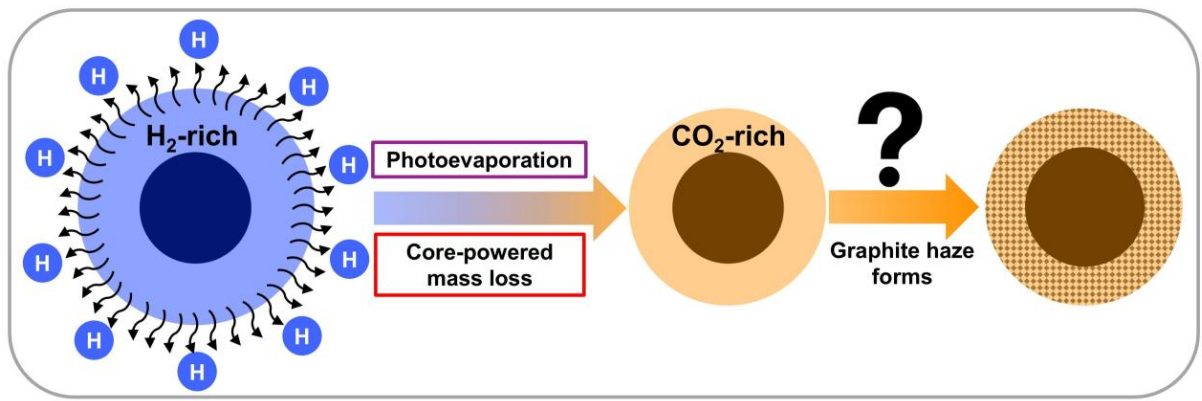
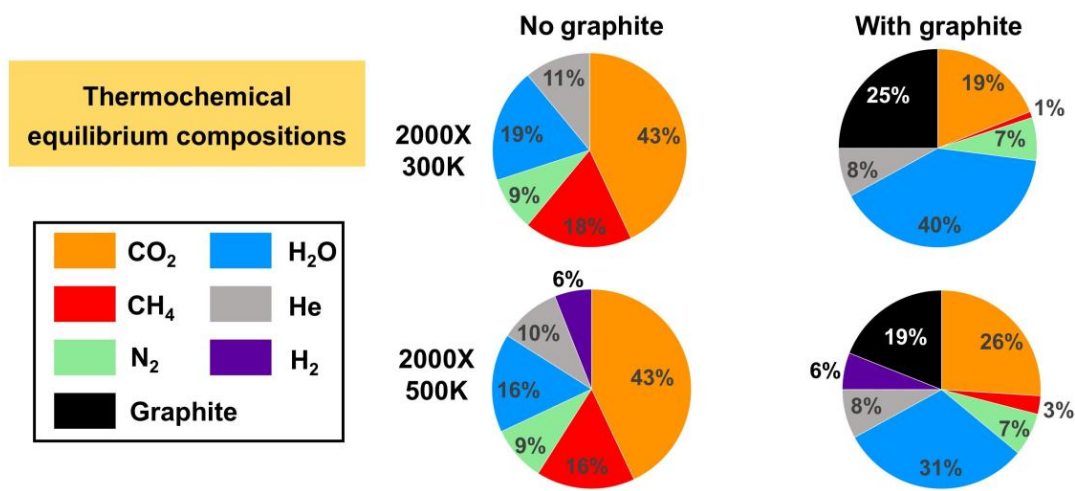

**Figure 2.** The upper panel illustrates the mass loss processes and potential formation of graphite in sub-Neptune atmospheres within the graphite-stability regime. The lower panel presents the atmospheric compositions calculated from the thermochemical equilibrium model without and with graphite.

### 2.1.2. Experimental Procedure

Detailed descriptions of the experimental procedures can be found in previous studies (C. He et al. 2018b; C. He et al. 2017; C. He et al. 2022; S. M. Hörst et al. 2018). Gas mixtures, excluding water vapour, were premixed in a stainless-steel cylinder with high-purity gases purchased from Airgas ($CO_2$ 99.999%, $CH_4$ 99.999%, $N_2$ 99.9997%, He 99.9995%, and $H_2$ 99.9999%). The water vapour was introduced from high-performance liquid-chromatography water (Fisher Chemical) at the required temperature to maintain the aimed mixing ratio, which was maintained by a dry ice, methanol and water cold bath. The total gas flow rate is 10 standard cubic centimeters per minute (sccm), and the pressure in the reaction chamber is held at 4.9 mbar at 300 K (5.3 mbar at 500 K). A heating coil was employed to heat the gas mixture (including the water vapour) to the experimental temperature (300 and 500 K). After flowing the heated gas mixtures through a reaction chamber for 4 hours, we examined the chamber with microscopy and did not observe the production of any solid particles (>300 nm) at either temperature. Then we exposed the heated gas mixture to an alternating current (AC) glow discharge (cold plasma) in the reaction chamber, facilitating chemical processes in the system. With the energy input, we noticed the formation of solid particles in less than 1 hour. To collect a sufficient amount of solid sample for further analysis, we let the experiments run continuously for ~100 hours.

During the experiments, we monitored the gas composition using a residual gas analyzer (RGA, Stanford Research Systems, RGA300, a quadrupole mass spectrometer). After the experiments, the chamber was sealed and transferred to a dry (< 0.1 ppm $H_2O$), oxygen-free (< 0.1 ppm $O_2$), $N_2$ glovebox (Inert Technology Inc., I-lab 2GB) for sample collection. The collected solid samples were weighed for production rate calculations, and were kept in the glovebox until further analysis to avoid contamination from Earth's atmosphere and light sources.

### 2.2. Gas Phase Product Analysis and Data Processing

The mass spectra were recorded for the gas mixtures before and during plasma exposure using a residual gas analyzer (RGA) to monitor the changes in gas composition resulting from the chemical reactions induced by the plasma discharge. The RGA system was linked to the chamber through a 30 cm stainless-steel tube. The gas molecules were ionized with a 70 eV electron impact source. The measurements were performed at a mass resolution of 1 amu with the mass/charge range (m/z) from 1-100 to detect and quantify the gas-phase products. Prior to plasma discharge, the pre-discharge background mass spectra of the initial gas mixture were acquired as a 30-scan average, establishing a control reference for subsequent reaction product analysis. After turning on the AC plasma discharge, the mass spectra averaged over 300 scans were acquired for the gas mixtures, enabling identification of reaction products generated through plasma-driven chemical reactions.

The RGA's unit resolution makes it challenging to distinguish species by exact mass due to overlapping signals. To identify and quantify the newly formed species, we needed to deconvolve the mass spectra. In theory, using the RGA's calibration fragmentation patterns, we can determine the species' mixing ratios by solving a linear combination of their relative abundance and fragmentation patterns (T. Gautier et al. 2020). However, the fragmentation pattern of the same molecule varies among different instruments, and calibration data of various molecules relevant to our samples on the RGA were not available. We used calibration data from the National Institute of Standard and Technology (NIST) mass spectral library (except for the main initial gases $N_2$ and $CO_2$, which we measured with the RGA) and adapted a Monte Carlo based approach to vary the peak intensity of individual fragment ions for each species. This approach has been used for analyzing both lab RGA data (J. Bourgalais et al. 2020; C. He et al. 2022) and Ion and Neutral Mass Spectrometer (INMS) data (J. Serigano et al. 2020; J. Serigano et al. 2022) from the Cassini mission, enabling the resolution of overlapping peaks in the mass spectra and the determination of the relative abundances. In this study, we allowed the fragment ion intensities varying within ±40% to account for uncertainties on the mass-dependent gas transfer efficiency, geometry of ion source and detection efficiency by the detector. We allow the algorithm to run until it collects 3,000 data points that meet the tolerance criterion. We included 49 species (composed of C, H, N, and O elements) in the database for the fitting. The selection of these 49 species is based on the chemistry of the gas mixture and the signals of the mass spectra. Since the signals above 47 amu were near the noise level, the base peaks of the 49 molecules are all below 47 amu.

To ensure the comparability of data across different experimental runs and to account for variability introduced by both the instrument and experimental conditions, a normalization process was applied. For this purpose, helium (He) was chosen as the reference gas because helium is inert gas and should remain unchanged during the experiment. The He+ ion has a stable mass signal at m/z 4, which does not overlap with signals from other species. Therefore, this mass signal is an ideal reference to normalize the mass spectra before and during plasma irradiation. Normalizing the mass spectra to the intensity of the He peak eliminates potential variation from instrumental or experimental variations, such as fluctuations in detector sensitivity or changes in experimental conditions. This normalization effectively aligned the mass spectra from different experimental runs, providing a consistent reference across all measurements.

### 2.3. Solid Phase Product Analysis and Data Processing

For the collected solid haze samples, we determined the density of exoplanet haze using a method as described in previous studies (e.g., C. He et al. 2024). We measured the sample's mass using a high-precision analytical balance and its volume using a gas pycnometer (AccuPyc II 1340, Micrometrics). Then, the sample's density was determined by dividing the mass by the volume.

The chemical composition of the haze particles was measured using a Fourier transform infrared (FTIR) spectrometer and very high-resolution mass spectrometer. We obtained the spectral properties of the haze particles with FTIR (Vertex 70v, Bruker Optics) from 4000 to 500 cm$^{-1}$ (2.5-20 μm) to identify the functional groups in the samples. The transmittance spectra of the haze samples were measured using the potassium bromide (KBr) pellet method following the procedure described previously (C. He et al. 2024). The measurement was performed under vacuum (below 0.2 mbar) at room temperature (294 K) using a DLaTGS detector and a KBr beam splitter to cover 4000 to 500 cm$^{-1}$.

We also employed very high-resolution mass spectrometry (HRMS, LTQ-Orbitrap XL, Thermo Fisher Scientific) to study the composition of the haze samples. Its high resolving power (>10$^5$) and excellent mass accuracy (±2 ppm) allows for assignment of molecular formula to each mass peak. We followed the methods used in previous studies (S. E. Moran et al. 2022; S. E. Moran et al. 2020; Z. Yang et al. 2025). Each sample was dissolved in methanol (CH$_3$OH) at a concentration of 1 mg/mL. After sonification for 1 hour and centrifugation at 10,000 rpm for 10 minutes, the soluble portion of the sample was diluted again in CH$_3$OH at 1 mg/mL. The diluted soluble fraction was injected into the Orbitrap with electrospray ionization (ESI) for mass spectra measurements in positive and negative ionization modes. The high-resolution allows us to assign unique molecular formulas to single charged ions. Based on the identified molecular formulas (C$_x$H$_y$N$_z$O$_t$), we calculated the number of double bond equivalents (DBE), defined as the number of rings plus double bonds in a neutral molecule:

$$DBE = \#C - \frac{(\#H - \#N)}{2} + 1 \qquad (1)$$

Where *#C*, *#H*, and *#N* refer to the number of carbon, hydrogen and nitrogen atoms in each identified molecular formula, respectively. Note that oxygen does not affect to the DBE of the molecules.

## 3. Results and Discussion
### 3.1. Mass Spectra of Gas Phase Products and Deconvolution Results

We present the gas phase mass spectrometry results here, which shows chemical evolution of the gas-phase species under plasma discharge. Figure 3 shows the mass spectral deconvolution results using our Monte Carlo approach for the two experiments at 300 K and 500 K. Black outline bars represent the normalized mass spectrum of the plasma discharge experiments. Colored segments represent the contribution of each species to the mass channel as calculated by the model. The deconvolution method minimizes the sum of squared residuals between the model predictions and the data, which is conceptually similar to least squares fitting in two dimensions, thereby yielding an optimal solution. Note that the blank black bars are either at background noise level or do not match with any stable gas species.

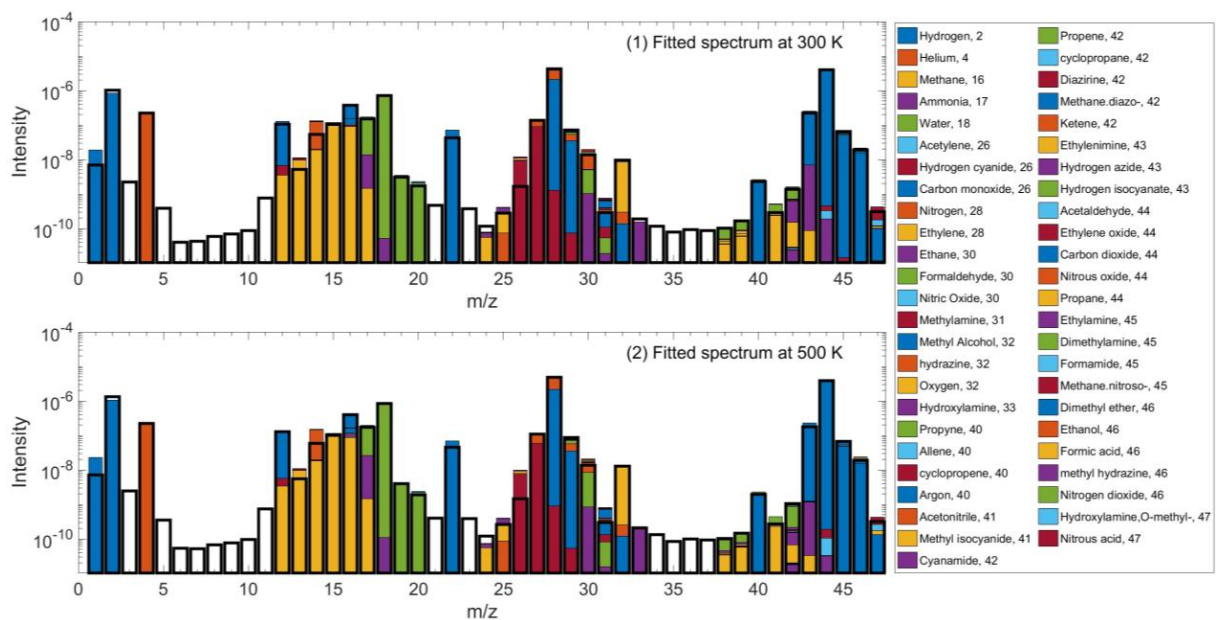

**Figure 3.** Mass spectral deconvolution results using our Monte Carlo approach for the normalized mass spectrum in the two experiments. Black outline bars represent the normalized mass spectra obtained from the plasma discharge experiments. Colored segments represent the contribution of each species to the mass channel as calculated by the model. The species in the legend are ordered by their respective molecular weights.

By comparing the deconvolution results with the initial gas composition, we calculated the abundance change of initial gas molecules and identified newly formed gas products. The result is summarized in Table 1, showing that the same gaseous products were generated at both temperatures. The fitted relative abundance are the average values of 3000 best-fitted data points after billions of simulations. The species presented in the table have relative abundance above the empirically defined threshold of $1.0 \times 10^{-10}$, and species with lower relative abundance are excluded to reduce errors from random fluctuations in data processing. This threshold exceeds the system's noise floor (~$6.5 \times 10^{-11}$), ensuring retained species have relative abundance distinct from background fluctuations and minimizing noise-induced uncertainty.

The initial gas mixtures in both experiments consisted of $CO_2$, $CH_4$, $N_2$, He, $H_2$, and $H_2O$ as determined by thermochemical equilibrium calculations for high-metallicity atmospheres. After plasma irradiation, significant changes in molecular composition were observed, with notable consumption of the initial gases and the emergence of a variety of new chemical species. The dissociation of $CO_2$, $CH_4$ and $N_2$ played a central role in driving these reactions, serving as the source of carbon, oxygen, hydrogen and nitrogen and leading to the formation of a diverse array of hydrocarbons, oxygenated organics, and nitrogen-containing molecules. The major products include $H_2$, $H_2O$, CO, $NH_3$, HCN, $N_2O$, and $CH_2O$. The declining trends for the initial gases and the resulting products are similar for the experiments at 300 K and 500 K. Helium is an inert gas and its relative abundance remains unchanged during the experiments. The major gas products are $H_2$, $H_2O$, and CO. $H_2$ is the recombination product of hydrogen atoms dissociated from $CH_4$. The reaction between $CO_2$ and $CH_4$ produces CO and $H_2O$. Note that the increase of $H_2$ and $H_2O$ does not mean that they did not participate the reaction in the system, only representing the net change of dissociation and production. The next major product is HCN, probably formed from the reaction of $CH_4$ and $N_2$ as occurs in Titan's atmosphere (S. M. Hörst 2017; A. Marten et al. 2002). The rest of the gas products are at relatively lower abundance, about $10^{-10}$-$10^{-9}$ levels. The molecular size of these detected gas products is up to 3 heavy atoms (C, N, or O), with molecular mass up to 47 amu. Heavier molecules may form in the system, but their abundance is below the detection limit of the RGA.

**Table 1.** Deconvolution results showing the relative abundance of gas phase species at 2000× metallicity for 300 K and 500 K conditions. The uncertainties represent the margins of error corresponding to the 95% confidence intervals for the 3,000 best-fit data points. Organics refers to gaseous products containing at least C and H, with the potential to also include O and N. Total denotes all the newly generated gas products.

| 2000× Metallicity at 300K | | | 2000× Metallicity at 500K | | |
|---|---|---|---|---|---|
| Species | Relative abundance | Uncertainties | Species | Relative abundance | Uncertainties |
| $CO_2$ | ↓$1.38 \times 10^{-6}$ | ±$4.84 \times 10^{-9}$ | $CO_2$ | ↓$1.19 \times 10^{-6}$ | ±$4.54 \times 10^{-9}$ |
| $N_2$ | ↓$4.32 \times 10^{-7}$ | ±$4.15 \times 10^{-10}$ | $N_2$ | ↓$2.36 \times 10^{-7}$ | ±$4.58 \times 10^{-10}$ |
| $CH_4$ | ↓$5.85 \times 10^{-8}$ | ±$1.21 \times 10^{-10}$ | $CH_4$ | ↓$5.64 \times 10^{-8}$ | ±$1.23 \times 10^{-10}$ |
| He | – | – | He | – | – |
| $H_2$ | ↑$5.74 \times 10^{-7}$ | ±$9.62 \times 10^{-10}$ | $H_2$ | ↑$7.26 \times 10^{-7}$ | ±$1.06 \times 10^{-9}$ |
| $H_2O$ | ↑$7.36 \times 10^{-8}$ | ±$1.04 \times 10^{-9}$ | $H_2O$ | ↑$2.11 \times 10^{-7}$ | ±$1.09 \times 10^{-9}$ |
| CO | ↑$1.07 \times 10^{-6}$ | ±$8.88 \times 10^{-9}$ | CO | ↑$1.09 \times 10^{-6}$ | ±$1.23 \times 10^{-8}$ |
| HCN | ↑$2.15 \times 10^{-8}$ | ±$1.16 \times 10^{-10}$ | HCN | ↑$1.52 \times 10^{-8}$ | ±$1.26 \times 10^{-10}$ |
| $N_2O$ | ↑$7.69 \times 10^{-9}$ | ±$5.92 \times 10^{-10}$ | $N_2O$ | ↑$7.52 \times 10^{-9}$ | ±$6.28 \times 10^{-10}$ |
| $NH_3$ | ↑$4.21 \times 10^{-9}$ | ±$2.38 \times 10^{-10}$ | $NH_3$ | ↑$8.85 \times 10^{-9}$ | ±$3.64 \times 10^{-10}$ |
| $CH_2O$ | ↑$1.64 \times 10^{-9}$ | ±$7.34 \times 10^{-11}$ | $CH_2O$ | ↑$3.19 \times 10^{-9}$ | ±$1.04 \times 10^{-10}$ |
| $HN_3$ | ↑$1.39 \times 10^{-9}$ | ±$1.26 \times 10^{-10}$ | $HN_3$ | ↑$2.33 \times 10^{-10}$ | ±$3.93 \times 10^{-11}$ |
| $C_2H_6O$ | ↑$8.09 \times 10^{-10}$ | ±$3.21 \times 10^{-11}$ | $C_2H_6O$ | ↑$1.12 \times 10^{-9}$ | ±$3.90 \times 10^{-11}$ |
| $NO_2$ | ↑$7.05 \times 10^{-10}$ | ±$5.72 \times 10^{-11}$ | $NO_2$ | ↑$4.72 \times 10^{-10}$ | ±$4.70 \times 10^{-11}$ |
| $HNO_2$ | ↑$6.81 \times 10^{-10}$ | ±$3.80 \times 10^{-11}$ | $HNO_2$ | ↑$4.37 \times 10^{-10}$ | ±$3.16 \times 10^{-11}$ |
| $C_2H_4$ | ↑$6.00 \times 10^{-10}$ | ±$1.43 \times 10^{-11}$ | $C_2H_4$ | ↑$5.94 \times 10^{-10}$ | ±$1.64 \times 10^{-11}$ |
| $C_2H_6$ | ↑$5.89 \times 10^{-10}$ | ±$1.74 \times 10^{-11}$ | $C_2H_6$ | ↑$4.86 \times 10^{-10}$ | ±$1.71 \times 10^{-11}$ |
| $C_2H_7N$ | ↑$4.79 \times 10^{-10}$ | ±$2.30 \times 10^{-11}$ | $C_2H_7N$ | ↑$3.80 \times 10^{-10}$ | ±$1.99 \times 10^{-11}$ |
| $CH_2O_2$ | ↑$2.96 \times 10^{-10}$ | ±$2.95 \times 10^{-11}$ | $CH_2O_2$ | ↑$7.41 \times 10^{-10}$ | ±$5.16 \times 10^{-11}$ |
| $CH_3NO$ | ↑$2.49 \times 10^{-10}$ | ±$3.95 \times 10^{-11}$ | $CH_3NO$ | ↑$4.09 \times 10^{-10}$ | ±$4.81 \times 10^{-11}$ |
| Organics | $2.64 \times 10^{-8}$ | | Organics | $2.22 \times 10^{-8}$ | |
| Total gases | $1.11 \times 10^{-6}$ | | Total gases | $1.15 \times 10^{-6}$ | |
| ↓: Decrease; ↑: Increase; –, no change because of Helium used as the reference. | | | | | |

The newly formed gas species demonstrate the formation of various hydrocarbons ($C_2H_4$ and $C_2H_6$) and nitrogen- and oxygen-containing compounds. Nitrogen-containing compounds include organic (HCN, $C_2H_7N$, and $CH_3NO$) and inorganic molecules ($N_2O$, $NH_3$, $NO_2$, $HNO_2$, and $HN_3$), while oxygenated species contain several oxides and organic molecules ($CH_2O$, $C_2H_6O$, $CH_2O_2$, and $CH_3NO$). Among these gas products, hydrogen cyanide, ammonia, and formaldehyde (HCN, $NH_3$, and $CH_2O$) are key precursors for organic haze formation as demonstrated in previous studies (e.g., A. Bar‑Nun & S. Chang 1983; C. He et al. 2018a; C. He et al. 2012; C. He et al. 2022; C. He & M. A.

Smith 2014a, 2014b). The experimental temperature does not change the composition of the new gas species but **may** affect their abundance.

### 3.2. Solid Phase Products Analysis Results
### 3.2.1. Production Rate and Density of Haze Particles

The plasma initiates the dissociation/ionization of the initial gases, producing various products; the smaller, volatile molecules stay in the gas phase, while the larger, refractory ones end up in the solid particles. For the collected haze particles, the production rates of the two experiments (dividing collected mass by the reaction time) are 2.8 mg hr$^{-1}$ and 1.2 mg hr$^{-1}$ for the 300 K and 500 K experiments, respectively. Note that the calculated production rates reflect the lower limit in the experiments, because it is impossible to fully collect all the generated solid products in practice. These production rates are lower than our standard Titan experiment (7.4 mg hr$^{-1}$) or the water-rich exoplanet experiments (~10 mg hr$^{-1}$) using the same set-up (C. He et al. 2017; C. He et al. 2024), which is probably due to the more oxidized initial gas mixtures used in the current experiments. Comparing the two experiments in this study, the production rate is significantly impacted by the temperature, less than half the rate at the higher temperatures. This could be caused by different reaction pathways at the two temperatures, though it is more likely that the produced species are easier to condense and form solid particles at 300 K compared to 500 K. Due to this difference, we should expect the composition and physical properties of the resulting solid particles to be different.

The density of haze particles is an important property within exoplanet atmospheres, impacting aerosol microphysics processes such as coagulation, transport, and sedimentation. The measured densities of the collected solid haze particles were 1.46 and 1.50 g cm$^{-3}$ (measurement uncertainties < 1%) at 300 K and 500 K. These densities are higher than the density (~1.3 g cm$^{-3}$) of haze analogs formed in water-rich atmospheres (C. He et al. 2024). A higher haze mass density could alter the vertical distribution of particles, as denser particles would more efficiently sediment to deeper layers of the atmosphere, thereby affecting the observed spectra and their interpretation. Thus, these haze density measurements provide reliable parameters for observational modeling of relevant exoplanets, facilitating more accurate simulations and analyses. The density of haze particles is determined by their chemical compositions. Compared to water-rich atmospheres, the composition of haze particles formed in CO$_2$-rich atmospheres is different, leading to variations in density. The 500 K haze's density is slightly higher than the density at 300 K because the higher temperature increases the likelihood of large molecule condensation, leading to larger molecular sizes. We further investigate the composition of the solid products using FTIR and HRMS.

### 3.2.2. FTIR Spectra of Haze Particles

Figure 4 displays the transmittance spectra of our 300 K (blue line) and 500 K (red line) exoplanet haze particles, across the 4000-500 cm$^{-1}$ range, with functional groups labeled near the absorption features. The spectra reveal the characteristic absorption features of various functional groups within the haze particles, such as O–H, N–H, C–H, C≡N, –N=C=N–, C=O, C=N, C=C, N–O, C–O and C–N (D. Lin-Vien et al. 1991), indicating a complex composition of the haze particles. The identified functional groups are similar to those found in water-rich atmospheres (C. He et al. 2024) with different intensities, reflecting variations in their relative concentration and the different molecular compositions. Compared to 300K, the absorption in the 3400-3100 cm$^{-1}$ (2.94-3.23 μm) range at 500K is more pronounced, accompanied by a red shift, indicating a higher concentration of N-H groups in the 500 K haze sample.

The absorption features for carbodiimide (N=C=N) at 2100 cm$^{-1}$ (4.76 μm) and double bonds (C=O, C=N, and C=C) at 1690-1640 cm$^{-1}$ (5.92-6.10 μm) are also more prominent for the 500 K haze sample, suggesting a higher degree of unsaturation. The compositional difference suggests the distinct formation pathways at different experimental conditions. The following high-resolution mass spectra can provide additional information regarding their composition.

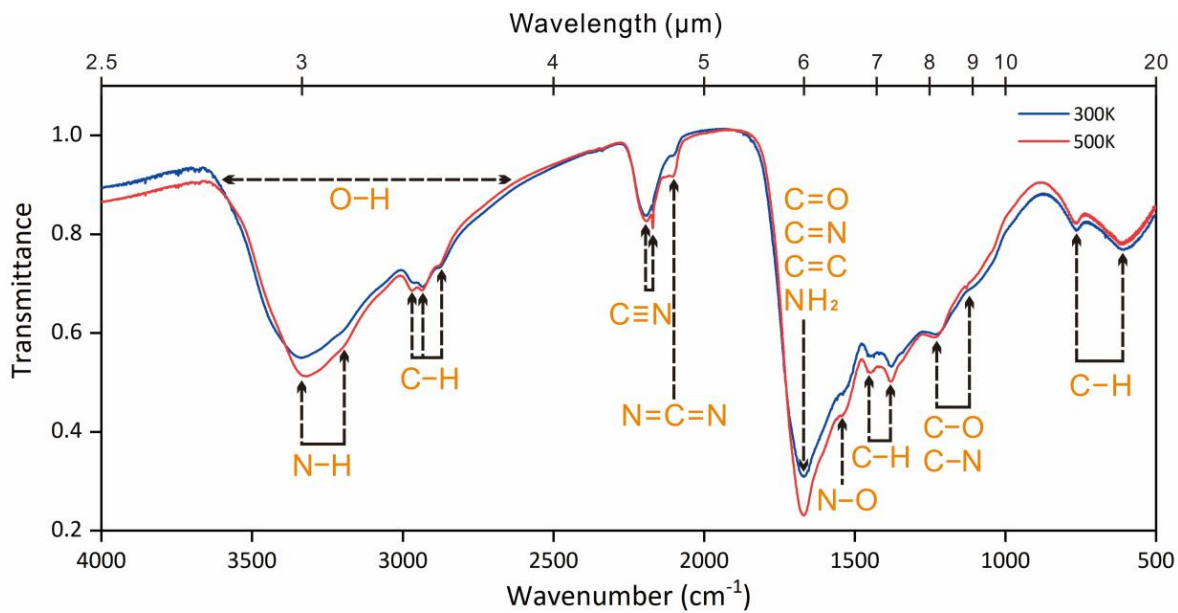

**Figure 4.** Transmittance spectra of two exoplanet haze particles formed in CO$_2$-rich gas mixtures at 300 and 500 K from 4000 to 500 cm$^{-1}$ to show the absorption features of various functional groups, which are labeled in the figure near their features.

### 3.2.3. Very High-Resolution Mass Spectrometry of Haze Particles

Figure 5 shows the very high-resolution mass spectrometry data of the solid products, revealing over thousands of mass spectral peaks in both positive and negative ionization modes. This indicates that the chemical composition of the haze particles formed in CO$_2$-dominant exoplanets is highly complex. The mass spectra demonstrate a repeating pattern of ~14 amu across both the positive and negative ions, a signature of an organic homologous series with increasing -CH$_2$ structural units. The Orbitrap uses an ESI technique, pronated ions which are detected as [M+H]$^+$ or [M-H]$^-$ when operating in positive or negative ionization mode, respectively (V. Vuitton et al. 2021). For the same sample, different species appear in positive or negative ionization modes due to their distinct acid-base properties of the molecule causing differing ionization efficiencies in either mode. We identified 2564 molecular formulas in positive ionization mode and 3546 in negative ionization mode for the 300 K sample, among them only 286 formulas appear in both positive and negative ionization modes. This is similar for the 500 K sample (196 repeat among 2142 positive ions and 3859 negative ions), indicating that most species appear only in positive or negative ionization modes. Using the ESI ionization technique, basic molecular species (like amines and amides) are preferentially protonated and detected in positive mode, while acidic species (like carboxylic acid) tend to lose a proton and be detected as negative ions. Some species with both acidic or basic groups can form positive and negative ions and be detected in both modes. In our study, only less than 5% of detected formulas show up in both modes, so obtaining the mass spectra in both modes provides a more complete understanding of the composition. Although there are more species identified in the negative ionization modes, their intensities are systematically lower, which is typical of such samples (S. E. Moran et al. 2020).

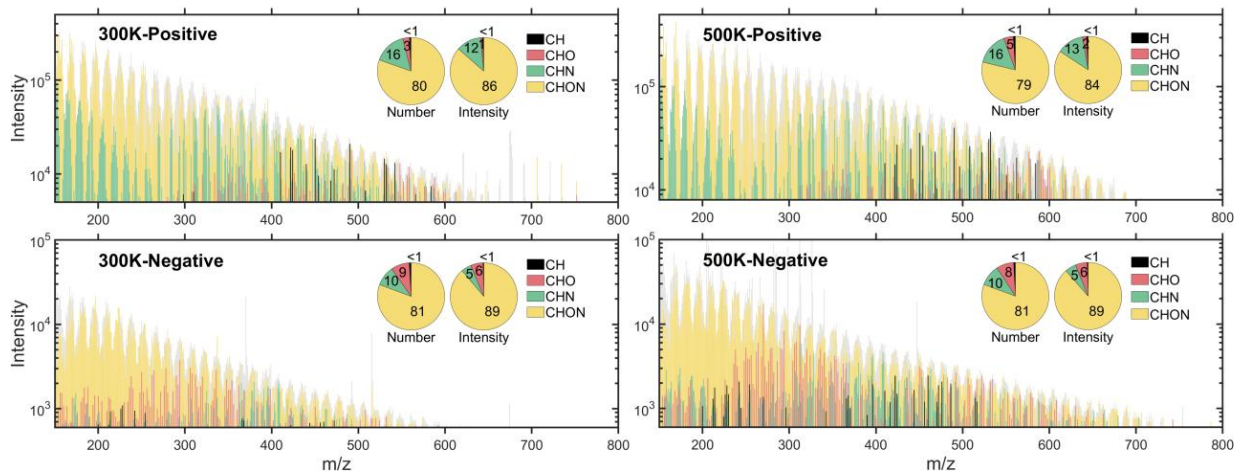

**Figure 5.** Positive and negative ionization mode mass spectra of the two haze particles from m/z 150 to 800 at 300 or 500 K. The yellow corresponds to the CHON subgroup while black to the CH subgroup, red to the CHO subgroup, green to the CHN subgroup, and the gray represents data that did not match any molecular formulas within the allowed mass accuracy (<2 ppm). The pie charts in the upper-right corner of each panel show the percentage of assigned molecular formulas in each subgroup (i.e., the number of molecular formulas matched within each subgroup) and the peak abundance percentage (i.e., the sum of the peak intensities of the matched compounds within each subgroup) for each organic compound subgroup.

The formulas composed of different elements are plotted in Figure 5 (black for the CH subgroup, red for CHO, green for CHN, and yellow for CHON). It is obvious the yellow peaks are dominant in all four mass spectra. To quantify different subgroups, we performed a statistical analysis on the variety and abundance of each subgroup. The number and intensity of the corresponding subgroups, represented as percentages in the respective ionization modes, are illustrated in the pie charts. The CHNO subgroup exhibits significantly higher values compared to the other subgroups in both samples. The CHN and CHO subgroups follow, with the CH subgroup being the lowest (less than 1%), suggesting that complex reactions happening in the $CO_2$-rich atmospheres lead to significant incorporation of both nitrogen and oxygen into the haze particles.

Based on the identified molecular formulas and their intensities, we calculated the average molecular formulas and average molecular weight for the two solid samples. Since the overlap of molecular formulas detected in both positive and negative ionization modes for the same sample is relatively small (less than 5%), the molecular formulas from both modes were combined and the abundance-weighted average molecular formulas were determined. The average molecular formulas for the sample at 300 K are $C_{13.3648}H_{20.2537}O_{2.6403}N_{4.5530}$ with a molecular weight of 286.78 amu, and $C_{15.2053}H_{23.8887}O_{2.8566}N_{4.8815}$ with a molecular weight of 320.59 amu for the 500 K samples. The average molecular size in the 500 K sample is larger, consistent with the higher haze density observed for this sample. Additionally, the higher nitrogen content in the average molecular formula in the 500 K sample further supports the previous FTIR result, showing more nitrogen-containing bonds present in this sample.

Furthermore, to visualize the composition of the analyzed sample, the Van Krevelen diagrams (D. W. Van Krevelen & K. Te Nijenhuis 2009) in Figure 6 display the data distribution in two dimensions (H/C vs. O/C and H/C vs. N/C).

The figure also shows the distribution of double bond equivalents (DBE), calculated using Equation (1), with values ranging from 0 to 30, reflecting the molecular complexity in the haze samples. Comparing the Van Krevelen diagrams, it is evident that temperature has little impact on the overall trends and does not significantly alter the general distribution trends of molecules. However, the compounds formed at both 300 K and 500 K show obvious differences between the positive and negative ionization modes, with the negative ionized species occupying a wider area in the plot. As overlayed on the 300 K H/C vs. O/C plot in Figure 6, the species detected in negative ionization mode are more scattered across the phase space, expanding to region (b) and region (c). Region (b) represents compounds with lower H/C and O/C ratios, where the degree of unsaturation decreases with increasing O/C ratio. In region (c), with higher H/C and O/C ratios, the species detected in the negative ionization modes show a lower degree of unsaturation. Regardless of H/C ratio, species with oxygen incorporation tend to form negative ions and are therefore detected in the negative ionization mode. These oxygen-containing species have lower degree of unsaturation, indicating that oxygen incorporates into the compounds as O-H bonds (alcohols and polyols like sugars) rather than C=O double bonds. The positive ionization mode detected more compounds in region (a), with high H/C ration and low O/C ratio. These compounds show lower unsaturation degree, likely saturated hydrocarbons with limited oxygen and nitrogen incorporation because they are more easily ionized with positive ESI mode. Similar to the H/C vs. O/C diagrams, the H/C vs. N/C diagrams also demonstrate a complexity of molecules with more diverse negatively ionized species. As shown in the lower-left panels of Figure 6, a greater number of species are observed in regions (e), (d), and (f) in the negative ionization mode. In region (e), both the H/C and N/C ratios are low, but the degree of unsaturation in relatively high (DBE>15), indicating higher number of double and triple bonds between carbon and carbon or carbon and nitrogen in these molecules. In both region (d) and region (f), the higher N/C ratio does not increase the degree of unsaturation, suggesting that nitrogen atoms are incorporated as amine groups that could be formed from nucleophilic substitution by $NH_3$. It is not surprising to see different distributions for molecular formulas detected in positive and negative ionization modes (Figure 6), because formulas detected in positive and negative ionization modes are largely different.

The Van Krevelen diagrams demonstrate radial patterns of species distribution in certain locations on the O/C and N/C vs. H/C phase space. Taking the 500 K samples as a reference (four panels on the right of Figure 6), we labeled these specific locations in the figure with numerical symbols (①, ②, ③, and ④). For the formulas identified in positive mode, the data points start from the labeled location ① (H/C = 2 and O/C = 0 or N/C = 0) and extend outward, which indicate that these species grow from saturated hydrocarbons $(C_2H_4)_X$ and form oxygenated/nitrogenated molecules with the increase of O/C or N/C ratio. There are some unsaturated species near location ② (H/C = 1) with very low O/C and N/C ratios, mainly unsaturated hydrocarbon like aromatics or polyyne; as N/C ratio increases, nitrogen incorporates into these unsaturated hydrocarbons. Moving on to the negative mode, the detected species are more complex and their distribution on the Van Krevelen diagrams are more scattered. Since the upper and lower panels represent the same set of molecules, we can identify two radiating centers, locations ③ and ④, despite the centers having different horizontal positions on the diagrams. These centers correspond to $(CH_2O)_y$ at location ③ and $(HCN)_z$ at location ④, hence why location ③ is further right on the O/C axis and location ④ is further right on the N/C axis. Radiating outward from these locations suggests the molecules may be built from these basic structural units.

The species between these two locations are likely the products of combination of these two units; the species close to location ③ have higher O/C ratio with more $CH_2O$ units while the ones near location ④ may be generated from more HCN units. The pattern shown in the Van Krevelen diagrams (Figure 6) further supports the hypothesis that the haze particles are produced from the reactive gas precursors like $C_2H_4$, $CH_2O$, HCN, which are all detected in the gas phase products.

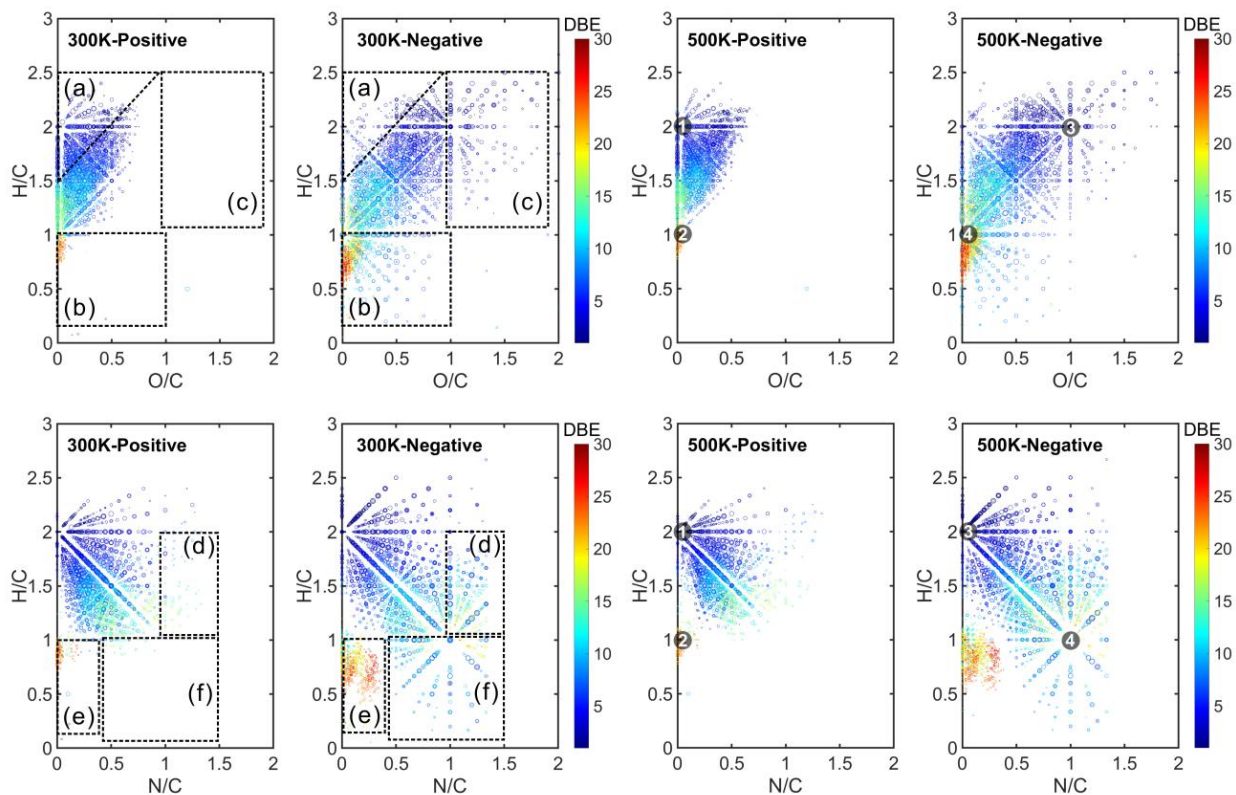

**Figure 6.** Van Krevelen diagrams (H/C vs. O/C and H/C vs. N/C) of the positive and negative ionization modes at the two experimental conditions. The color code represents the molecules' double bond equivalents (DBE), and the symbol size is proportional to the peak intensity of the very high-resolution mass spectrometry. The dashed lines mark regions representing differences caused by the ion patterns. The labeled numerical symbols (①, ②, ③, and ④) indicate the centers of the radiating pattern.

### 3.3. Discussion

Under the low pressure and low temperature conditions of our experiments, graphite haze does not form in $CO_2$-rich atmospheres. Instead, a variety of organic haze compounds are generated. Graphite consists of $sp^2$-hybridized carbon atoms arranged in a two-dimensional hexagonal lattice. In order to form graphite, the carbon atoms have to be freed from $CO_2$ or $CH_4$ in the gas mixture, which requires breaking all the bonds in $CO_2$ or $CH_4$ molecules and forming new bonds between the carbon atoms in a highly ordered crystalline structure. Since graphite has a low free energy, the formation of graphite is thermodynamically favorable. Our results suggest that under our specific experimental conditions, the reaction is kinetically inhibited because the formation pathways require high activation energy. Previous studies have shown that the synthetic processes for graphite usually require temperatures exceeding 2,100 ºC (H. Marsh & F. Rodríguez-Reinoso 2006). The energy source used in the experiment, provides sufficient energy to directly dissociate stable molecules (such as $CH_4$, $CO_2$, and $N_2$) to form radicals or ions. However, the energy density, which is approximately 170 W/m$^2$ (C. He et al. 2018a), is not high enough to simultaneously break all bonds in $CH_4$ or $CO_2$, since the average residual time of the gas is only about 2 seconds in our flowing reaction system. Therefore, it is difficult to meet the necessary energy requirements for graphite formation in our system. In contrast, organic compounds can form through free radical and/or ion chemistry without the need to break all bonds in the original carbon-containing gases. More specifically, under the energy input from the plasma, the dissociation and ionization of $CH_4$ and $CO_2$ both provide carbon sources to the formation of various hydrocarbons and oxygenated compounds, such as $C_2H_4$, $C_2H_6$, CO, and $CH_2O$ among many other products in the gas phase. Similarly, $N_2$ dissociates to produce nitrogen free radicals, which then react to form nitrogen-containing compounds, such as $NH_3$ and HCN. The gas products can further react to produce large, complex molecules that condense to haze particles. For instance, the identified gas molecules HCN, $NH_3$, and $CH_2O$ are key precursors for organic haze formation in Titan and exoplanet atmospheres, as suggested in previous studies (e.g., C. He et al. 2018a; C. He et al. 2022; E. Sciamma-O'Brien et al. 2010; B. N. Tran et al. 2008). This means that the $CO_2$-rich atmospheres that we expect on sub-Neptunes closer to the radius gap could host organic hazes instead of graphite in their upper atmospheres that are observable with transit spectroscopy. While in the deep atmospheres, where the pressure and temperature increase, the condition might meet for the formation of graphite, but it would not be detectable with current and near future telescopes unless the small graphite particles are lofted into the observable regions.

Many sub-Neptunes have been discovered, and the TESS mission continues to find more of this type of exoplanet. Among them, the exoplanets previously mentioned (such as GJ 436 b, GJ 1214 b, GJ 3470 b, TOI-776 c, and TOI-270 d) have already been observed by JWST and their spectra suggest the presence of hazes in their atmospheres. Our results indicate that these hazes are composed mainly of organics rather than graphite. The organic hazes have very different optical properties from graphite, leading to distinct observational effects (P. Gao et al. 2021). Characterizing the optical properties of the resulting organics hazes is important to analyze and interpret observational data of those sub-Neptunes, which is currently under investigation. This work is the first exoplanet lab experiment to explore the formation of organic haze in such atmospheres, advancing our understanding of atmospheric chemistry in this type of exoplanet. Future work on the optical properties of these organic hazes will help understand their impact on the observed spectra and enable the identification of key spectroscopic features in hazy atmospheres.

## 4. Conclusions

In summary, we conducted laboratory experiments to simulate $CO_2$-rich sub-Neptune atmospheres within the graphite-stability regime. We found that the haze composition was dominated by organic compounds in our experimental conditions. The mass spectra analysis of gas products reveals that plasma-driven chemistry in $CO_2$-rich sub-Neptune atmospheres leads to the transformation of initial gases ($CO_2$, $CH_4$, $N_2$, $H_2O$, and $H_2$) into a wide variety of hydrocarbons, oxygen- and nitrogen-containing species. Further analysis of the solid-phase products provides deeper insights into the composition and physical properties of haze particles. The FTIR and HRMS results reveal that the organic hazes contain various functional groups (O–H, N–H, C–H, C≡N, –N=C=N–, C=O, C=N, C=C, N–O, C–O and C–N) and great diversity of molecular formulas with different degree of unsaturation. These results improve our understanding of atmospheric chemistry in $CO_2$-rich sub-Neptune atmospheres. The organic hazes formed in these atmospheres are expected to impact the observed spectra of these exoplanets differently from graphite hazes. The investigation of their optical properties is underway to help understand the impacts and guide observations.

## Acknowledgments

The authors gratefully acknowledge the supports from the National Natural Science Foundation of China (42475132) and the U.S. National Science Foundation (2206245). S.E.M. is supported by NASA through the NASA Hubble Fellowship grant HST-HF2-51563 awarded by the Space Telescope Science Institute, which is operated by the Association of Universities for Research in Astronomy, Inc., for NASA, under contract NAS5-26555. V.Vuitton acknowledges support from the French National Research Agency in the framework of the "Investissements d'Avenir" program (ANR-15-IDEX-02), through the funding of the Origin of Life project of the Université Grenoble Alpes and the French Space Agency (CNES) under their "Exobiologie, Exoplanètes et Protection Planétaire" program.

# References

Acheson, E. G. 1896, in, ed. U. S. P. Office (US)
Bar-Nun, A., & Chang, S. 1983, Journal of Geophysical Research: Oceans, 88, 6662
Batalha, N. M., et al. 2013, The Astrophysical Journal Supplement Series, 204, 24
Beatty, T. G., et al. 2024, The Astrophysical Journal Letters, 970, L10
Berger, T. A., et al. 2018, The Astrophysical Journal, 866, 99
Berger, T. A., et al. 2020, The Astronomical Journal, 160, 108
Bonfanti, A., et al. 2024, Astronomy & Astrophysics, 682, A66
Bourgalais, J., et al. 2020, Scientific Reports, 10, 10009
Bower, D. J., et al. 2025, arXiv preprint arXiv:250700499
Chen, H., & Rogers, L. A. 2016, The Astrophysical Journal, 831, 180
Cloutier, R., & Menou, K. 2020, The Astronomical Journal, 159, 211
Davenport, B., et al. 2025, arXiv preprint arXiv:250101498
Fressin, F., et al. 2013, The Astrophysical Journal, 766, 81
Fulton, B. J., & Petigura, E. A. 2018, The Astronomical Journal, 156, 264
Fulton, B. J., et al. 2017, The Astronomical Journal, 154, 109
Gao, P., et al. 2021, in (Wiley Online Library)
Gautier, T., et al. 2020, Rapid Communications in Mass Spectrometry, 34, e8684
Ginzburg, S., Schlichting, H. E., & Sari, R. e. 2018, Monthly Notices of the Royal Astronomical Society, 476, 759
Gressier, A., et al. 2024, The Astrophysical Journal Letters, 975, L10
Gupta, A., & Schlichting, H. E. 2019, Monthly Notices of the Royal Astronomical Society, 487, 24
Gupta, A., & Schlichting, H. E. 2020, Monthly Notices of the Royal Astronomical Society, 493, 792
He, C., et al. 2018a, ACS Earth and Space Chemistry, 3, 39
He, C., et al. 2018b, The Astronomical Journal, 156, 38
He, C., et al. 2017, Astrophysical Journal Letters, 841
He, C., Lin, G., & Smith, M. A. 2012, Icarus, 220, 627
He, C., et al. 2024, Nature Astronomy, 8, 182
He, C., et al. 2022, ACS Earth and Space Chemistry, 6, 2295
He, C., & Smith, M. A. 2014a, Icarus, 243, 31
He, C., & Smith, M. A. 2014b, Icarus, 238, 86
Hörst, S. M. 2017, Journal of Geophysical Research: Planets, 122, 432
Hörst, S. M., et al. 2018, Nature Astronomy, 2, 303
Howard, A. W., et al. 2012, The Astrophysical Journal Supplement Series, 201, 15
Jin, S., et al. 2014, The Astrophysical Journal, 795, 65
Kempton, E. M.-R., et al. 2023, Nature, 620, 67
Lee, E. J., & Connors, N. J. 2021, The Astrophysical Journal, 908, 32
Lin-Vien, D., et al. 1991, The handbook of infrared and Raman characteristic frequencies of organic molecules (Elsevier)
Lopez, E. D., & Fortney, J. J. 2013, The Astrophysical Journal, 776, 2
Luque, R., & Pallé, E. 2022, Science, 377, 1211
Malsky, I., et al. 2025, The Astronomical Journal, 169, 221
Marsh, H., & Rodríguez-Reinoso, F. 2006, in Activated Carbon, eds. H. Marsh, & F. Rodríguez-Reinoso (Oxford: Elsevier Science Ltd), 454
Marten, A., et al. 2002, Icarus, 158, 532
Mayor, M., et al. 2011, arXiv preprint arXiv:11092497
Mbarek, R., & Kempton, E. M.-R. 2016, The Astrophysical Journal, 827, 121
Modirrousta-Galian, D., Locci, D., & Micela, G. 2020, The Astrophysical Journal, 891, 158
Moran, S. E., et al. 2022, Journal of Geophysical Research: Planets, 127, e2021JE006984
Moran, S. E., et al. 2020, The Planetary Science Journal, 1, 17
Mordasini, C. 2020, Astronomy & Astrophysics, 638, A52
Moses, J. I., et al. 2013, The Astrophysical Journal, 777, 34
Mukherjee, S., et al. 2025, The Astrophysical Journal Letters, 982, L39
Ohno, K., et al. 2025, The Astrophysical Journal Letters, 979, L7
Owen, J. E. 2019, Annual Review of Earth and Planetary Sciences, 47, 67
Owen, J. E., & Schlichting, H. E. 2024, Monthly Notices of the Royal Astronomical Society, 528, 1615

Owen, J. E., & Wu, Y. 2013, The Astrophysical Journal, 775, 105
Rogers, J. G., & Owen, J. E. 2021, Monthly Notices of the Royal Astronomical Society, 503, 1526
Schlawin, E., et al. 2024, The Astrophysical Journal Letters, 974, L33
Sciamma-O'Brien, E., et al. 2010, Icarus, 209, 704
Sciamma-O'Brien, E., Ricketts, C. L., & Salama, F. 2014, Icarus, 243, 325
Serigano, J., et al. 2020, Journal of Geophysical Research: Planets, 125, e2020JE006427
Serigano, J., et al. 2022, Journal of Geophysical Research: Planets, 127, e2022JE007238
Stevenson, K. B., et al. 2010, Nature, 464, 1161
Teske, J., et al. 2025, The Astronomical Journal, 169, 249
Tran, B. N., et al. 2008, Icarus, 193, 224
Van Eylen, V., et al. 2018, Monthly Notices of the Royal Astronomical Society, 479, 4786
Van Krevelen, D. W., & Te Nijenhuis, K. 2009, Properties of polymers: their correlation with chemical structure; their numerical estimation and prediction from additive group contributions (Elsevier)
Vuitton, V., et al. 2021, The Planetary Science Journal, 2, 2
Woitke, P., et al. 2021, Astronomy & Astrophysics, 646, A43
Wolf, E. T. 2017, The Astrophysical Journal Letters, 839, L1
Yang, J., & Hu, R. 2024, The Astrophysical Journal Letters, 971, L48
Yang, Z., et al. 2025, The Planetary Science Journal, 6, 47